\documentclass[3p,times]{elsarticle}

\usepackage{amssymb}
\usepackage{wrapfig}
\usepackage{graphicx}
\usepackage{hyperref}
\usepackage{amsmath}
\usepackage{placeins}
\usepackage{tikz}
\usepackage[normalem]{ulem} 
\usepackage[makeroom]{cancel} 
\usetikzlibrary{shapes.geometric, arrows}
\usetikzlibrary{positioning}
\usepackage{caption}
\usepackage{subcaption}
\usepackage{booktabs}
\usepackage{float}

\newcommand{\X}{\mathbf{X}}

\begin{document}

\begin{frontmatter}

\title{Time-Optimal Operation of a Load-Hoisting Gantry Crane}

\author[mysecondaryaddress]{Eric Mountain}

\author[mysecondaryaddress]{Tarunraj Singh}
\cortext[mycorrespondingauthor]{Corresponding author}
\ead{tsingh@buffalo.edu}

\address[mysecondaryaddress]{Department of Mechanical and Aerospace Engineering, University at Buffalo (SUNY),\\ Buffalo, NY 14260-4400, USA}

\begin{abstract}
This paper addresses the problem of designing time-optimal control profiles for point-to-point control of a gantry crane moving in a two dimensional plane. It is assumed that the hoisting motor completes the hoisting maneuver at a constant rate and completes its transition in the same time that it takes for the cart to reach its terminal position. This results in a linear time-varying model and a closed form solution to the time-optimal control problem is shown to be parameterized with Bessel functions. A comprehensive analysis of the structure of the time-optimal control profile is studied by examining the switching function which illustrates the mechanism of introduction and decimation of switches in the bang-off-bang control profile. The variation of the number of switches in the optimal control profile is presented and a non-intuitive control profile structure is noted, one which initiates with a stationary cart, while the hoisting cable length is changed prior to the initiation of motion of the cart. To address the issue of uncertainties in the initial cable length, a model with the sensitivity of the system states with respect to the initial cable length is used to augment the system model and is used for the design of robust time-optimal controllers. Experimental results validate the time-optimal and robust time-optimal control profiles. 
\end{abstract}


\end{frontmatter}

\section{Introduction}\label{sec:intro}
Gantry cranes serve as essential machinery for heavy materials handling in construction sites, manufacturing plants, and shipyards around the world. Modern innovations in controls theory have been applied to improve crane automation, making them safer, more efficient, and more cost effective. As a result, industries and researchers have sought to quantify the advantages and drawbacks of  crane automation, particularly with regards to their safety and economic impact.

Sisson~\cite{sisson2011}, a senior port planner with expertise in port operation, posits a 40\% reduction in injuries with full crane automation.
In addition, the authors of the report ``Container Port Automation: Impacts and Implications'' published at the International Transport Forum in 2021 \cite{OECD}, analyzed data collected from 63 countries and summarized their findings about the state of automation in ports. The report highlights a statement from Asea Brown Boveri (ABB), a leading crane manufacturer, which claims a 45-55\% reduction in labor time per crane due to automation. Regarding labor costs, Oliveira and Varela \cite{oliveira_varela_2017} show a 33\% labor cost reduction per automated crane. In another study, Rosenfeld and Shapira~\cite{rosenfeld1998automation} assessed 11 sites to gauge the impact of retrofitting tower cranes with control systems to improve crew productivity, and enhance safety. Their conservative estimate indicated a return on investment within one year. It can be readily concluded that the productivity gains and safety improvements stem from better control of crane operations.

There exists a long and extensive history of research focused on the motion control of cranes.
Abdel-Rahman et al.~\cite{abdel2003dynamics} present in their review article details of lumped and distributed modeling approaches for cranes and generate reduced order models and approximate solutions to these models. They also present a thorough review of existing control methodologies. More recently, the book {\it Dynamics and Control of Industrial Cranes}~\cite{hong2019dynamics} reviews various types of cranes and presents lumped and distributed parameter models for these cranes. It reviews open-loop and closed-loop control strategies that are prevalent in industrial applications.
The review article and the book are testimony to the interest in the development of control strategies for cranes given their ubiquitous presence in construction, ports, offshore rigs, etc. 

Mojallizadeh et al.~\cite{mojallizadeh2023modeling} present a comprehensive review of modeling and control for overhead cranes, where a large fraction of papers reviewed used a lumped mass representation of the system dynamics. Rather than modeling the cable of the crane and a massless rigid link, it has also been modeled as a distributed parameter system.
d’Andrea-Novel et al.~\cite{d1994feedback} represent the crane cable by the wave equation and design an asymptotically stable collocated controller. Alli and Singh~\cite{alli1999passive} proposed a passivity based control for an overhead crane where the crane cable is modeled as a distributed parameter system.

There have been numerous studies addressing the problem of concurrent cart and hoisting motion which makes the model nonlinear~\cite{auernig1987time, singhose2000effects, ramli2020efficient}. Reference shaping has been used extensively to attenuate residual vibrations of the suspended mass~\cite{singhose2000effects,masoud2006graphical, ur2022input}. Singhose et al.~\cite{singhose2000effects} consider a simplified model of a gantry crane where the hoisting velocity is considered a control variable which is not shaped since the cable is considered to be inflexible. Since the frequency of oscillation of the cable changes with hoisting, shapers are designed based on the initial length, the average of the initial and terminal frequencies, and the frequency corresponding to the average cable length. Results show that averaging resulted in better vibration attenuation relative to shapers based on initial frequency. Rehman et al. \cite{ur2022input} design an adaptive controller for an underactuated tower crane with hoisting and mass variations. A neural network was used to develop a nonlinear input-output mapping of the parameters and for design of a Zero Vibration Derivative shaper. Experimental results show that the shaper is capable of robustly improving payload lifting operations and reducing residual vibrations. These input shaping approaches target the residual vibrations and not the maneuver time.

Within the many studies analyzing the difficulties of simultaneous cart and hoisting motion, there has been extensive research in applications with constant hoisting velocity. Auernig and Troger~\cite{auernig1987time} determine time-optimal control strategies for a gantry crane with traversing motion in conjunction with constant hoisting motion. Different control profiles are analyzed for differing mechanical and electrical hardware applications, and simulations on five ship unloaders operating in European ports validate these findings. Furthermore, Abdel-Rahman and Nayfeh \cite{Rahman2002} study cable length manipulations and their effects on cargo pendulations in boom cranes. Numerous constant hoisting velocities were used to determine optimal hoisting speeds that reduce these pendulations. Elling and McClinton \cite{Elling1973} investigate the dynamic loading cranes placed on ships that are undergoing arbitrary prescribed motion. The hoisting velocity in this study is held constant for each excitation period.

It is not difficult to justify the need for minimum-time point-to-point maneuvering in cranes since that serves as a proxy for productivity. Kuntze and Strobel~\cite{kuntze1975contribution} presented a problem formulation for time-optimal control of a crane without hoisting, where they assume a three switch bang-bang control profile. Recently Stein and Singh~\cite{stein2023minimum} presented a comprehensive analysis of the change in the structure of the time-optimal control profile when the velocity of the cart is constrained to be positive. Auernig and Troger's~\cite{auernig1987time} defined new states related to the center of mass of the system, and arrived at a nonlinear set of equations which are used for the optimal control design. In contrast, in this work, a simplified linear time-varying model is used assuming that the velocity of the cart can be commanded.

This paper initiates with the articulation of the assumption and the development of a third order model for the gantry crane. This is followed by the formulation of the time-optimal control problem in Section~\ref{sec:3}, which develops the necessary conditions for optimality and synthesizes the closed form solutions representing the evolution of the costates and states. Section~\ref{sec:sw_tran} develops the constraints which lead to the addition or removal of switch time(s) in the bang-off-bang control profile. Details of the evolution of the time-optimal control profile as a function of the cart displacement is presented to illustrate that the number of switches in the optimal control is a function of the terminal displacement of the cart. The impact of varying hoisting distance on the switching profiles is visualized via three dimensional switching surfaces in conjunction with the maneuver time surface. To address the issues of uncertainty in initial conditions of the cable length, a state sensitivity based approach is used in Section \ref{sec:robcon} to design bang-off-bang control profiles which are robust to initial length uncertainties. Finally, in Section~\ref{sec:6} experimental results are presented to validate the numerical results and support the assumptions made in arriving at the system model.
\section{System Model}\label{sec:prob_st}

\begin{wrapfigure}[11]{r}{0.35\textwidth}
\vspace{-0.5in}
  \begin{center}
    \includegraphics[width=0.3\textwidth]{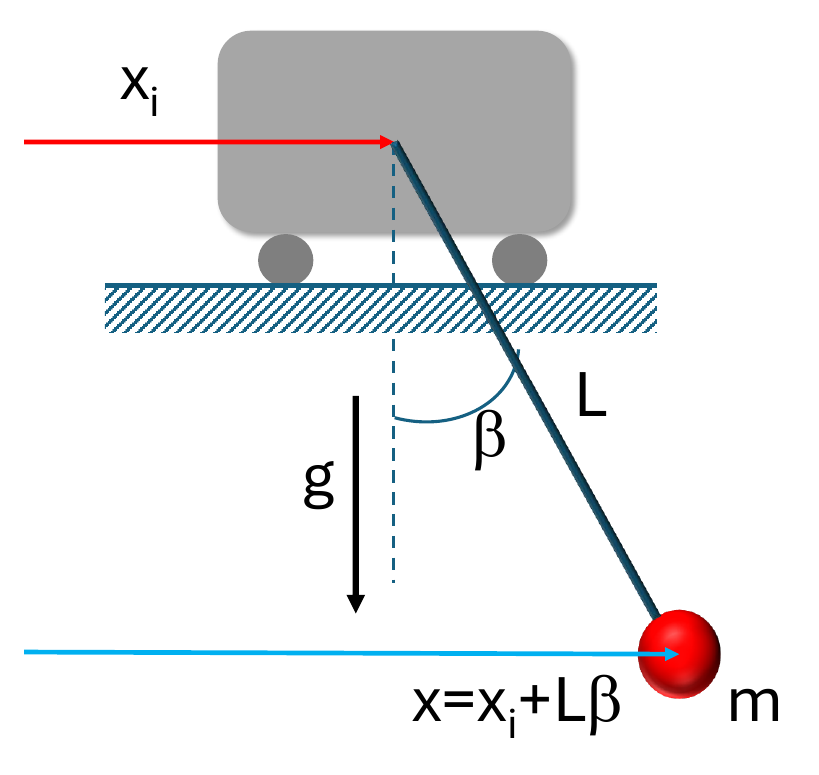}
  \end{center}
  \vspace{-0.3in}
  \caption{Gantry Crane System}
   \label{fig:01}
\end{wrapfigure}

The gantry crane setup includes a cart driven by a stepper motor which permits the commanding of the cart's velocity by assuming that the acceleration is zero.  The same assumption is made of the hoisting stepper motor, i.e., its velocity can be commanded. A schematic of the crane is shown in Fig.~\ref{fig:01} where a small angle displacement is assumed~\cite{Zhang.2014}~\cite{Chen.2016b} and where the hoisting motor is assumed to move with constant velocity and the cart velocity is the sole control input. The equations of motion can be expressed as: 

\begin{align}
    & mL^2\ddot{\beta}(t) + 2mL\dot{L}\dot{\beta(t)} + mgL\sin\left(\beta(t)\right) = 0\\
    & m\ddot{x}(t) - m\cancelto{0}{\ddot{x}_i(t)} + 2m\dot{L}\left(\frac{\dot{x}(t)-\dot{x}_i(t)}{L}\right) + mg\left(\frac{x(t)-x_i(t)}{L}\right) = 0\\
    & \leftrightarrow \color{black} \ddot{x}(t) + \frac{2\dot{L}}{L} \left( \dot{x} - \dot{x}_i \right) + \frac{g}{L}\left( x(t) - x_i(t)\right) = 0  \\
    & \dot{x}_i(t) = u(t) \label{eq:gantry} \\
    & \dot{L} = \frac{L_f-L_0}{t_f}
\end{align}

\noindent where $m$ is the payload mass, $L$ is the cable length, $\beta(t)$ is the swing angle, $g$ is gravity, $x_i$ is the cart displacement, $x$ is the payload displacement, $L_0$ is the initial cable length and $L_f$ is the terminal cable length. The velocity of the cart $u$ is considered as the input and is constrained to $0 \le u\le U_m$, where $U_m$ is the upper bound for the cart velocity. The assumption that the cart velocity can be used as a control input is based on papers which demonstrate velocity input control of industrial cranes~\cite{Sorensen.2007, Suksabai.2020, Knierim.2010, Wu.2021}. It is also assumed that the hoisting input $\dot{L}$ is a constant over the duration of the maneuver, which implies that the transition of the cable from its initial length to its terminal length is completed in the maneuver time $t_f$ that it takes for the cart to move from its initial location to its terminal location. This leads to a closed form representation of the cable length:
\begin{equation}
    L(t) = L_0 +\frac{L_f-L_0}{t_f} t, \: \: \: \forall \: \: \: 0\le t \le t_f.
\end{equation}
The final assumption is that $\dot{L}$ is very small and its impact on the dynamics can be ignored. The final time-varying state space model of the crane is:
\begin{equation}
\underset{\dot{\X}}{\underbrace{\begin{bmatrix}\dot{x}_1 \\ \dot{x}_{2} \\ \dot{x}_3 \end{bmatrix}}} = 
	\underset{A(t,t_f)}{\underbrace{\begin{bmatrix} 0 & 1 & 0 \\ -\frac{g t_f}{L_0 t_f + (L_f-L_0)t} & 0 & \frac{g t_f}{L_0 t_f + (L_f-L_0)t} \\ 0 & 0 & 0 \end{bmatrix}}} \underset{\X}{\underbrace{\begin{bmatrix} x_1 \\ x_2 \\ x_3 \end{bmatrix}}} + 
	\underset{B}{\underbrace{\begin{bmatrix} 0 \\ 0 \\ 1 \end{bmatrix}}} u,
\end{equation}
where $x_1$ is the payload position, $x_2$ is the payload velocity, and $x_3$ is the cart position.

\section{Time-Optimal Control}
\label{sec:3}

The time-optimal control problem for the rest-to-rest maneuver of the crane can be posed as:
\begin{subequations}
	\begin{align}
	& \mbox{min} ~~J = \int_{0}^{t_f} dt \\
	\mbox{subject to} & \notag \\ & \dot{\X} = A(t,t_f)\X+Bu \\
	& \X(0) = \begin{Bmatrix} 0 & 0 & 0 \end{Bmatrix}^T  \\
	&  \X(t_f) = \begin{Bmatrix} x_f & 0 & x_f \end{Bmatrix}^T \\
	& 0 \le u \le U_m \: \: \: \: \forall t.
	\end{align}
\end{subequations}
Defining the Hamiltonian\index{Hamiltonian} as:
\begin{equation}
\mathcal{H} = 1+\lambda^{T} \left( A(t,t_f)\X+Bu \right),
\end{equation}
the necessary conditions for optimality can be derived using calculus of variations,
resulting in the equations
\begin{subequations} \label{eq:topt_4}
	\begin{align}
	& \dot{\X} = \frac{\partial \mathcal{H}}{\partial \lambda} = A(t,t_f)\X+Bu \\
	& \dot{\bf{\lambda}} = -\frac{\partial \mathcal{H}}{\partial \X} = -A(t,t_f)^{T} \bf{\lambda} \label{eq:costates}\\
	& u = U_m \mathbf{H}(-\it{B}^T\bf{\lambda}) \label{eq:toptc}\\
	& \X(0) = 0 \mbox{ and } \X(t_f) = \begin{Bmatrix} x_f & 0 & x_f \end{Bmatrix}^T \\
	& \int_0^{t_f} \frac{\partial \mathcal{H}}{\partial t_f} \: dt = -\mathcal{H}(t_f) = -1 - \lambda_3(t_f) u(t_f) \label{eq:10e},
	\end{align}
\end{subequations}
where $\mathbf{H}$ is the Heaviside step function. Eq.~\eqref{eq:toptc} is derived using {\it Pontryagin's minimum principle (PMP)} which requires the optimal cart velocity to be bang-off-bang. 

The Hamiltonian is:
\begin{equation}
\mathcal{H} = 1+\lambda_1 x_2 + \lambda_2 \left( -\frac{g t_f}{L_0 t_f + (L_f-L_0)t} x_1 + \frac{g t_f}{L_0 t_f + (L_f-L_0)t} x_3\right) + \lambda_3 u
\end{equation}
and the resulting costate equations are:
\begin{align}
	\dot{\lambda}_1 &=-\frac{\partial \mathcal{H}}{\partial x_1} =  \lambda_2\frac{g t_f}{L_0 t_f + (L_f-L_0)t} \\
	\dot{\lambda}_2 &=-\frac{\partial \mathcal{H}}{\partial x_2} =  -\lambda_1 \label{eq:lam2}\\
	\dot{\lambda}_3 &=-\frac{\partial \mathcal{H}}{\partial x_3} =  -\lambda_2\frac{g t_f}{L_0 t_f + (L_f-L_0)t} \label{eq:lam3}
\end{align}

Consider the time derivative of Eq.~\eqref{eq:lam2}
\begin{equation}
    \ddot{\lambda}_2 =  -\dot{\lambda}_1 = -\lambda_2\frac{g t_f}{L_0 t_f + (L_f-L_0)t}
    \label{eq:L2}
\end{equation}
which can be represented in a standard second order Bessel equation form, whose solution is:
\begin{equation}
    \lambda_2 =  \sqrt{L_0 t_f + (L_f-L_0)t} \left(C_1 J_1 \left( \Phi(t) \right) + C_2 Y_1 \left( \Phi(t) \right)\right)
    \label{eq:lam2sol}
\end{equation}
where $\Phi(t) = 2 \sqrt{\frac{g t_f}{(L_f-L_0)^2} \left(L_0 t_f + (L_f-L_0)t\right)}$,
and $J_1$ and $Y_1$ are the Bessel function of the first and second kind of order 1 respectively. Substituting Eq. \eqref{eq:lam2sol} into Eq. \eqref{eq:lam3} and integrating leads to:
\begin{equation}
    \lambda_3(t) = \sqrt{gt_f}\operatorname{sgn}(L_f-L_0)(C_1J_0(\Phi(t))+C_2Y_0(\Phi(t))) + C_3
\end{equation}

Details of deriving $\lambda_3(t)$ can be found in \ref{appendix_B}. The constants $C_1, C_2$ and $C_3$ are solved for based on the switch times of the optimal bang-off-bang control. For instance, if the optimal control is parameterized as:
\begin{equation}
    u = V_m \left( 1 - \mathbf{H}(t-T_1) + \mathbf{H}(t-T_2) - \mathbf{H}(t-T_3)\right)
\end{equation}
where $T_1$ and $T_2$ are the switch times and $T_3$ is the maneuver time, the zero crossing constraints of the switching function are: 
\begin{align}
    \lambda_3(T_1)&=  \sqrt{gT_3}\operatorname{sgn}(L_f-L_0)\left(C_1 J_0 \left( \Phi(t=T_1) \right) + C_2 Y_0 \left( \Phi(t=T_1) \right)\right) + C_3 =0 \\
        \lambda_3(T_2) &=  \sqrt{gT_3}\operatorname{sgn}(L_f-L_0)\left(C_1 J_0 \left(\Phi(t=T_2)\right) + C_2 Y_0 \left( \Phi(t=T_2)\right)\right) + C_3 =0  
\end{align}
which require the constants $C_1$, $C_2$ and $C_3$ to lie in the null space of the $2 \times 3$ matrix $P$: 
\begin{equation}
    P = \begin{bmatrix}
       J_0 \left( \Phi(t=T_1)\right) & Y_0 \left( \Phi(t=T_1)\right) & \frac{1}{\sqrt{g T_3}\operatorname{sgn}(L_f-L_0)} \\
       J_0 \left( \Phi(t=T_2)\right) & Y_0 \left( \Phi(t=T_2)\right) & \frac{1}{\sqrt{gT_3}\operatorname{sgn}(L_f-L_0)}
    \end{bmatrix}.
\end{equation}
Since the null space only provides a vector, to scale the vector appropriately, Eq.~\eqref{eq:10e} is used as a constraint. 

With the knowledge that the optimal control is bang-off-bang, the integral of Eq.~\eqref{eq:gantry} which represents the cart displacement is a sum of delayed ramps. 
To arrive at a closed form expression for the motion of the suspended mass of gantry crane system, we first consider the solution of the equation:
\begin{equation}
    \ddot{x} + \frac{g t_f}{L_0 t_f + (L_f-L_0)t} x = \frac{g t_f}{L_0 t_f + (L_f-L_0)t} t
    \label{eq:22}
\end{equation}
which is the response to a ramp input ($x_i = x_3 = t$):
\begin{equation}
    x(t) =  t + \sqrt{L_0 t_f + (L_f-L_0)t}\left(K_1 J_1 \left( \Phi(t) \right) + K_2 Y_1 \left( \Phi(t) \right) \right) 
    \label{eq:disp_t}
\end{equation}
and 
\begin{align}
    \dot{x}(t) &=  1 + \frac{(L_f-L_0)}{2\sqrt{L_0 t_f + (L_f-L_0)t}}\left(K_1 J_1 \left( \left( \Phi(t) \right)\right) + K_2 Y_1 \left( \Phi(t)  \right) \right)+ \notag\\
    &  \frac{1}{2}\sqrt{L_0 t_f}\left(K_1 \left( J_0(\Phi(t)) -\frac{1}{\Phi(t)}J_1(\Phi(t))\right) + K_2 \left( Y_0(\Phi(t)) -\frac{1}{\Phi(t)}Y_1(\Phi(t))\right) \right)
    \label{eq:vec_t}
\end{align}
where $K_1$ and $K_2$ are determined based on the initial conditions $x(0)$ and $\dot{x}(0)$. Details of representing Eq.~\eqref{eq:22} in the standard Bessel equation form are provided in \ref{appendix_A}.

To determine the solution to the optimal control profile, we need to consider the response of the suspended mass when it is subject to a series of delayed ramps. First, we consider the response to the sum of a ramp and a delayed ramp input:
\begin{equation}
    u = t - (t-T) \mathbf{H}(t-T).
\end{equation}
The solution given by Eqs.~\eqref{eq:disp_t} and ~\eqref{eq:vec_t} is valid for $0\le t \le T$, at which time, a similar solution is generated with a different particular integral. It should be ensured that the states match at $t=T$. The solution over the time interval $T\le t \le t_f$ is given as:
\begin{equation}
    x(t) =  T + \sqrt{L_0 t_f + (L_f-L_0)(t-T)}\left(D_1 J_1 \left( \Phi(t-T) \right) + D_2 Y_1 \left( \Phi(t-T) \right) \right) 
    \label{eq:disp_t2}
\end{equation}
and 
\begin{align}
    &\dot{x}(t) =  0 + \frac{(L_f-L_0)}{2\sqrt{L_0 t_f + (L_f-L_0)(t-T)}}\left(D_1 J_1 \left( \left( \Phi(t-T) \right)\right) + D_2 Y_1 \left( \Phi(t-T)  \right) \right)+ \notag \\
    &\frac{1}{2}\sqrt{L_0 t_f}\left(D_1 \left( J_0(\Phi(t-T)) -\frac{1}{\Phi(t-T)}J_1(\Phi(t-T))\right) + D_2 \left( Y_0(\Phi(t-T)) -\frac{1}{\Phi(t-T)}Y_1(\Phi(t-T))\right) \right)
    \label{eq:vec_t2}
\end{align}
where $D_1$ and $D_2$ are integration constants. We require 
Equations~\ref{eq:disp_t} and \ref{eq:vec_t} at $t=T$ to be equal to 
Equations~\ref{eq:disp_t2} and \ref{eq:vec_t2} at their initial time $t=T$, which leads to the constraints:


\begin{equation}
\begin{aligned} 
&T + \sqrt{L_0 t_f + (L_f-L_0)T}
  \left(K_1 J_1(\Phi(T)) + K_2 Y_1(\Phi(T))\right)
= \\ 
&T + \sqrt{L_0 t_f}
  \left(D_1 J_1(\Phi(0)) + D_2 Y_1(\Phi(0))\right)
\end{aligned}
\end{equation}

or
\begin{equation}
    \sqrt{L_0 t_f + (L_f-L_0)T}\left(K_1 J_1 \left( \Phi(t) \right) + K_2 Y_1 \left( \Phi(t) \right) \right)  =  \sqrt{L_0 t_f}\left(D_1 J_1 \left( \Phi(0) \right) + D_2 Y_1 \left( \Phi(0) \right) \right) 
\end{equation}
and

\begin{equation}
\begin{aligned}
  &1 + \frac{(L_f-L_0)}{2\sqrt{L_0 t_f + (L_f-L_0)T}}
     \left(K_1 J_1(\Phi(T)) + K_2 Y_1(\Phi(T)) \right) \\
  &+ \frac{1}{2}\sqrt{L_0 t_f}
     \left(K_1 \left( J_0(\Phi(T)) -\frac{1}{\Phi(T)}J_1(\Phi(T))\right)
     + K_2 \left( Y_0(\Phi(T)) -\frac{1}{\Phi(T)}Y_1(\Phi(T))\right) \right) = \\
  & \frac{(L_f-L_0)}{2\sqrt{L_0 t_f}}
     \left(D_1 J_1(\Phi(0)) + D_2 Y_1(\Phi(0)) \right) \\
  &+ \frac{1}{2}\sqrt{L_0 t_f}
     \left(D_1 \left( J_0(\Phi(0)) -\frac{1}{\Phi(0)}J_1(\Phi(0))\right)
     + D_2 \left( Y_0(\Phi(0)) -\frac{1}{\Phi(0)}Y_1(\Phi(0))\right) \right)
\end{aligned}
\end{equation}

We have two equations in two unknowns which can be solved for $D_1$ and $D_2$. This process can be used to solve for any input profile which consists of sums of delayed ramps. 
This permits posing a constrained parameter optimization problem to solve for the switch times and the maneuver times of the optimal control profile.

For a two switch bang-off-bang control profile parameterized as:
\begin{equation}
    u = V_m \left( 1 - \mathbf{H}(t-T_1) + \mathbf{H}(t-T_2) - \mathbf{H}(t-T_3)\right),
\end{equation}
we solve the problem:
\begin{subequations}
\begin{align}
 & \mbox{min} ~~J = T_3 \\
\mbox{subject to} & \notag \\ 
 & x_1(0) = 0 \mbox{ and } x_1(T_3) = x_f \\
 & x_2(0) = 0 \mbox{ and } x_2(T_3) = 0 \\
 & x_3(0) = 0 \mbox{ and } x_3(T_3)= U_m (T_3+T_1-T_2) = x_f \\
  & 0 \le T_1 \le T_2 \le T_3 
\end{align}
\end{subequations}
where $x_1(T_3)$ and $x_2(T_3)$ are analytically represented by Bessel functions as described before. 

\begin{figure}[h]
\centering
\includegraphics[width=1.0\linewidth]{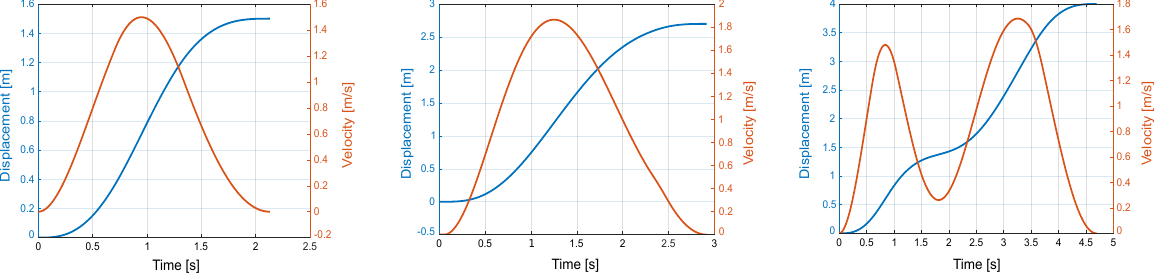}
\caption{Displacement and Velocity Profiles. (a) Two Switch Profile with $x_f$ = 1.5 m (b) Three Switch Profile with $x_f$ = 2.7 m (c) Four Switch Profile with $x_f$ = 4 m}\label{fig:states}
\end{figure}

The optimal state evolution for terminal displacement of $x_f = 1.5, 2.7$, and $4$ m are illustrated in Fig.~\ref{fig:states}


\begin{figure}[h]
\centering
\includegraphics[width=1.0\linewidth]{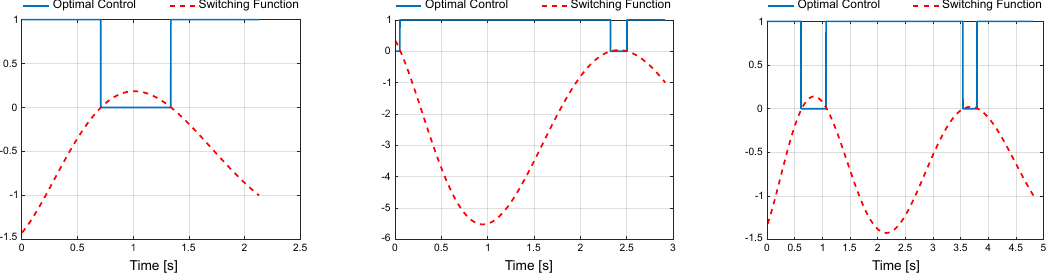}
\caption{Optimal Control and Switching Function Profiles. (a) Two Switch Profile with $x_f$ = 1.5 m (b) Three Switch Profile with $x_f$ = 2.7 m (c) Four Switch Profile with $x_f$ = 4 m}\label{fig:control}
\end{figure}

\noindent and the associated optimal control and switching functions are illustrated in Fig.~\ref{fig:control}.

As the final displacement increases from $x_f = 1.5$ m to $x_f = 2.7$ m, a non-intuitive time-optimal profile emerges, one which starts with a magnitude of zero for a while before the cart starts to move as illustrated in Fig.~\ref{fig:states}(b) and \ref{fig:control}(b)
where it can be seen that the pendular motion is not initiated at $t=0$ s, but at $t=0.05$ s, over which interval the hoisting mechanism has initiated the change in the hoisting length of the cable.



 
\section{Switching Profile Transition}
\label{sec:sw_tran}

\begin{figure}[h!]
\centering
\includegraphics[width=.90\linewidth]{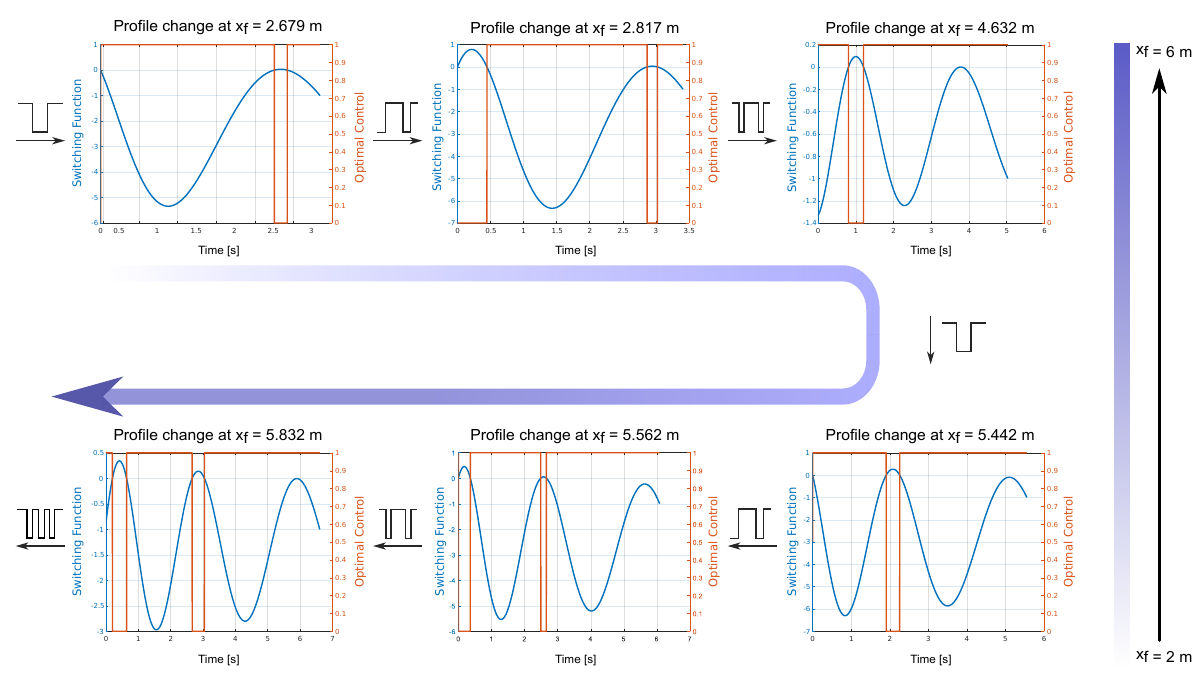}
\caption{Evolution of Control Profiles with Transitions}
\label{fig:transition_profiles}
\end{figure}

This section studies the variation in the structure of the optimal control as a function of variation of the terminal displacement of the cart while holding the total hoisting distance constant. The switching function for generating the time-optimal solution is $\lambda_3(t)$, which is characterized by Bessel functions of the first and second kind. 
With the knowledge that these functions become asymptotically harmonic as their argument increases, we have infinite zero crossings. This implies that the time-optimal control profile can have variations in the number of switches in the time-optimal control profiles. It should be noted that the switches can be birthed or decimated one at a time, which would occur at the start or termination of the maneuver, requiring 
\begin{equation}
\lambda_3(0) = 0, \mbox{ or } \lambda_3(t_f) = 0.
\end{equation}
One can also conceive of a scenario where two switches are introduced or destroyed concurrently. This can only happen at time instants ($\tau$) which lie between the initial time and terminal time of the maneuver which requires the following constraints to be satisfied:
\begin{align}
\lambda_3(t=\tau) = 0, \\
\dot{\lambda}_3(t=\tau) = 0,
\end{align}
which correspond to requiring the switching function and its slope being zero at the transition time $t=\tau$. A comprehensive analysis studying the variation in the structure of the time-optimal control profile is carried out for a specific change in hoisting distance where $L_0=1$ m and $L_f=3$ m. Fig.~\ref{fig:transition_profiles} illustrates the variation in the structure of the time-optimal control profile as the cart displacement varies from $x_f=2$ m to $x_f=6$ m. The time-optimal control profile initiates with a 2-switch control profile and the structure changes to ones with 3, 4, 2, 3, 4, and 6 switch profiles sequentially.


\begin{figure}
\centering
\includegraphics[width=.65\linewidth]{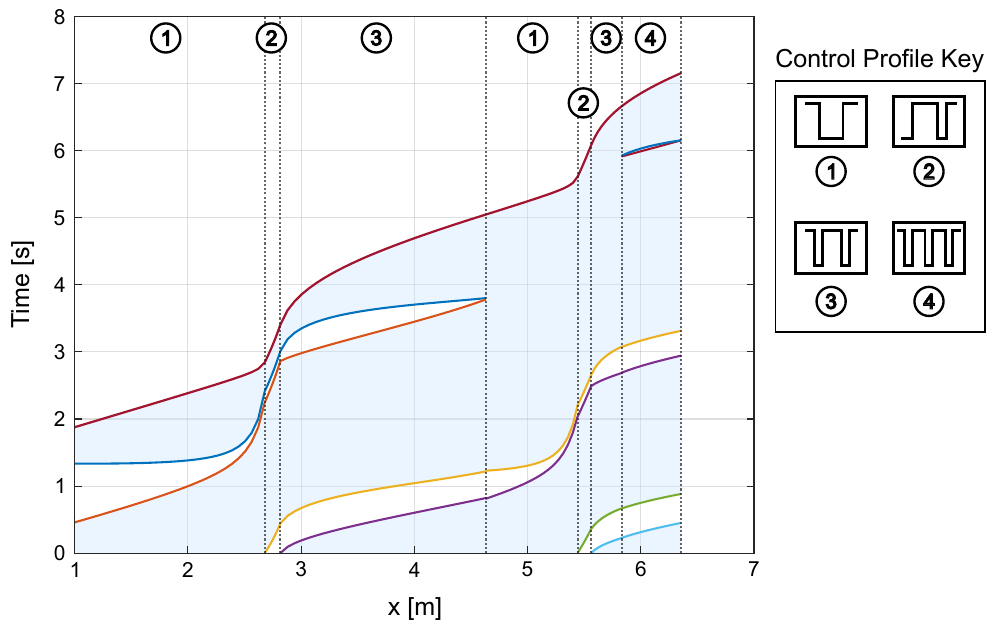}
\caption{Evolution of the Switch and Maneuver Times}\label{fig:switch_tf}
\end{figure}

The variation of the switch times and the maneuver time as a function of the terminal displacement is illustrated in Fig.~\ref{fig:switch_tf}. The shaded regions signify active cart control, while the blank regions represent pauses in the cart control. It should be noted that around displacements of 2.7 m and 2.8 m, single switches are introduced into the control profile. 

\begin{figure}
\centering
\includegraphics[width=.85\linewidth]{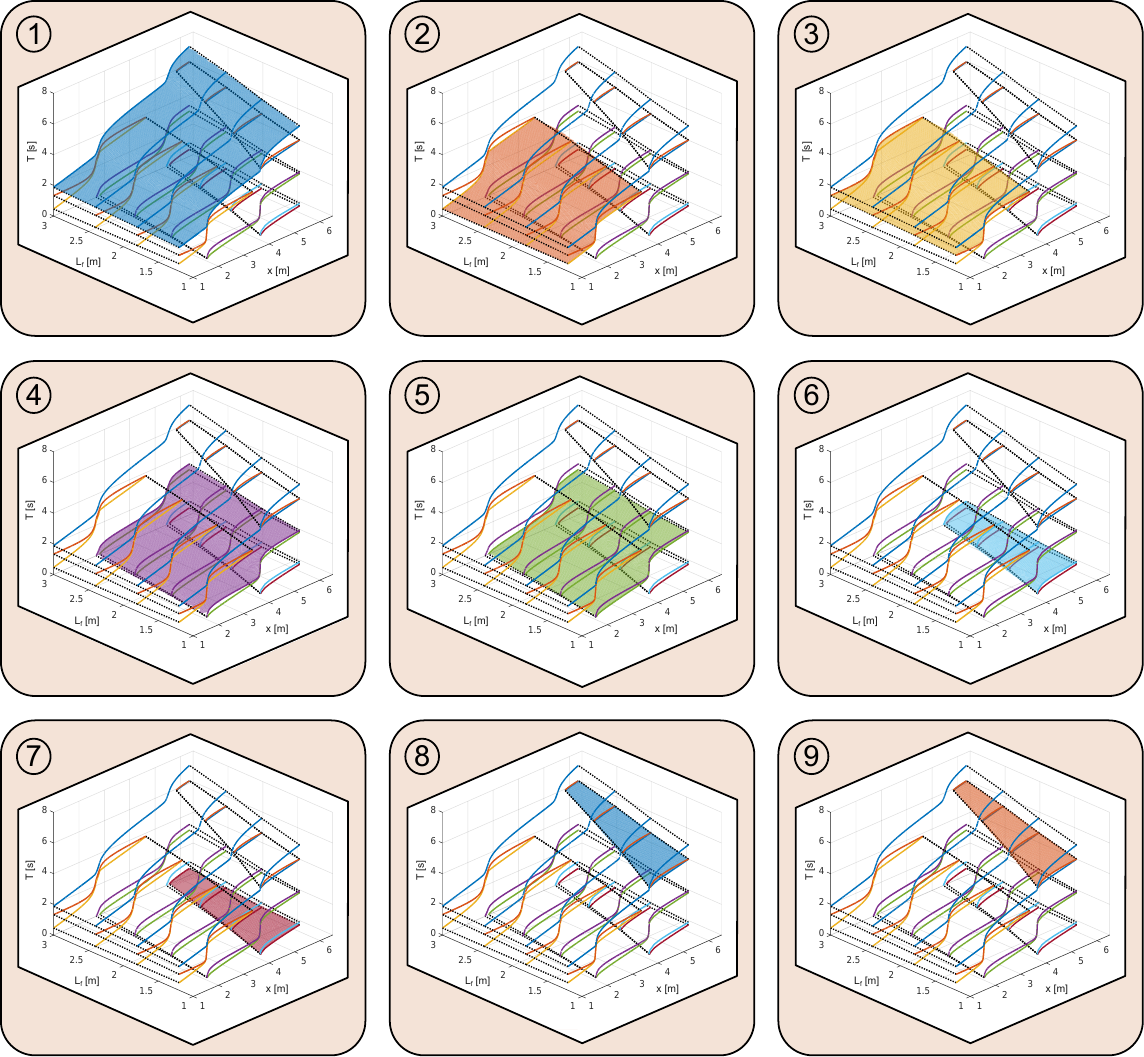}
\caption{Evolution Surfaces of the Switch and Maneuver Times}\label{fig:switch_surf}
\end{figure}

To study the transition in the structure of the control profile as a function of total hoisting distance and the cart motion, a three dimensional spectrum of surfaces are generated to illustrate the optimal switch time structure for various hoisting distances and cart motions. Fig.~\ref{fig:switch_surf} illustrates these surfaces where \textcircled{\raisebox{-0.2ex}{1}} represents the maneuver time variation, while \textcircled{\raisebox{-0.2ex}2} through \textcircled{\raisebox{-0.2ex}9} illustrate switch time evolution. The individual surfaces are plotted separately in the interest of clarity. The surfaces represent a pattern of smooth transitions in switch times as a function of the hoisting and cart displacements. They further visualize the birth and collapse of switches in this three dimensional space. Fig. \ref{fig:switch_tf} can now be interpreted as a single slice in the $T-x$ plane for an $L_f$ value of 3 m. It should be noted that surfaces \textcircled{\raisebox{-0.2ex}8} and \textcircled{\raisebox{-0.2ex}9} form a clamshell with an upper and lower surface, representing a birth and collapse of a bang-off-bang switch. This is the case for surfaces \textcircled{\raisebox{-0.2ex}6} and \textcircled{\raisebox{-0.3ex}7}, \textcircled{\raisebox{-0.2ex}4} and \textcircled{\raisebox{-0.2ex}5}, and \textcircled{\raisebox{-0.2ex}2} and \textcircled{\raisebox{-0.2ex}3}. The wireframe is provided to illustrate the switch time variation as illustrated in Fig.~\ref{fig:switch_tf} for specific values of terminal hoisting distance. Furthermore on the $x-L_f$ plane, solid lines indicate the variation of the hoisting distance as a function of cart displacement when new switches are introduced at the initial time, i.e., start of the maneuver.

\section{Robust Control}
\label{sec:robcon}
As has been evident, there are profound variations in the structure of the time-optimal control profiles for the point-to-point maneuvering of a gantry crane where in addition to the cart being displaced, the hoisting cable also changes length. This prompts the question of studying the impact of uncertainty in the estimated initial length of the hoisting cable and the synthesis of control profiles which are insensitive to errors in estimates of the initial length of the cable $L_0$.
In this work, we evaluate the sensitivity of the states of the system with respect to uncertainty in the initial cable length and force the state sensitivities with respect to the uncertain initial cable length to zero at the terminal time. The resulting augmented state space model is:
\begin{align}
\dot{x}_1(t) &= x_2(t)\\
\dot{x}_2(t) &= -\frac{g t_f}{L_0 t_f + (L_f-L_0)t} x_1(t) + \frac{g t_f}{L_0 t_f + (L_f-L_0)t} x_3(t)\\
\frac{d\dot{x}_1(t)}{d L_0} &= \frac{dx_2(t)}{d L_0}\\
\frac{d\dot{x}_2(t)}{dL_0} &= \frac{g t_f\left( t_f-t\right)}{\left(L_0 t_f + (L_f-L_0)t\right)^2} x_1(t) -\frac{g t_f}{L_0 t_f + (L_f-L_0)t}\frac{dx_1(t)}{dL_0} - \frac{g t_f\left( t_f-t\right)}{\left(L_0 t_f + (L_f-L_0)t\right)^2} x_3(t)\\
\dot{x}_3(t) &= u(t)\\
0 & \leq u \leq U_m.
\end{align}
and is subject to the initial and final conditions:
\begin{align}
x_1(0)&=x_2(0)=x_3(0)=0\\
\frac{dx_1}{d L_0}(0)&=\frac{dx_2}{d L_0}(0)=0 \\
x_1(t_f) &= x_3(t_f)= x_f, \\
x_2(t_f)&=\frac{dx_1}{d L_0}(t_f)=\frac{dx_2}{d L_0}(t_f)=0.
\end{align}

\begin{figure}
\centering
\includegraphics[width=.7\linewidth]{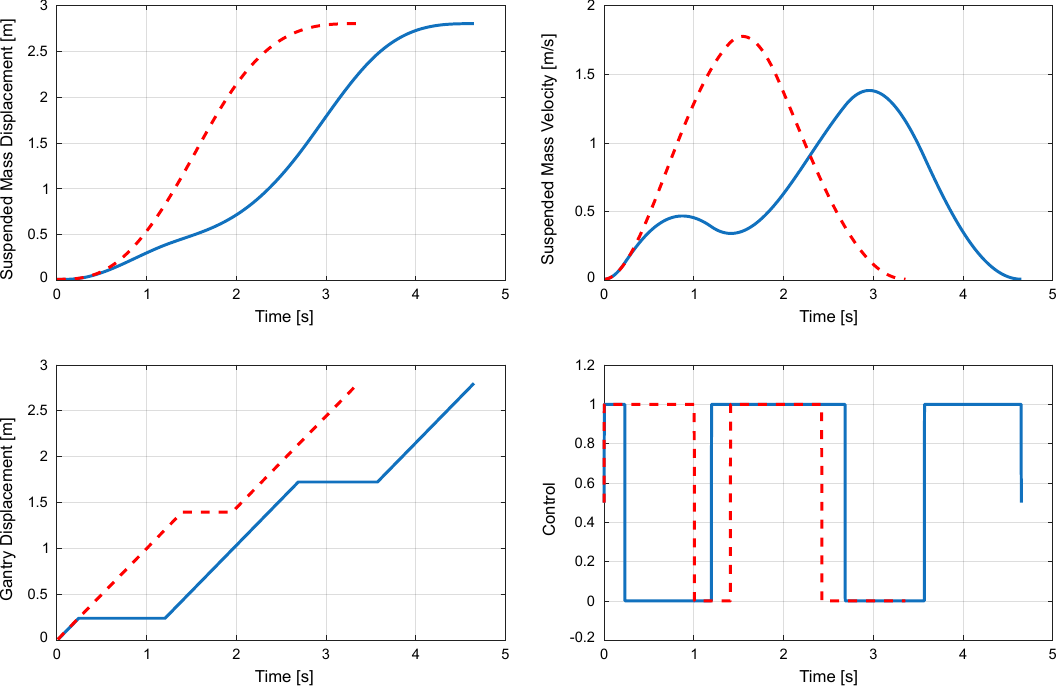}
\caption{Evolution of States and Control (Time-Optimal: Dashed Line, Robust Control: Solid Line)}\label{fig:robust_man}
\end{figure}

Fig.~\ref{fig:robust_man} illustrates the evolution of the states and the control for the time-optimal solution and the desensitized time-optimal solution where it is evident that there is a time penalty encumbered to desensitize the terminal residual energy to errors in the initial length of the pendulum.

\section{Experimental Results}
\label{sec:6}

\begin{figure}
\centering
\includegraphics[width=.6\linewidth]{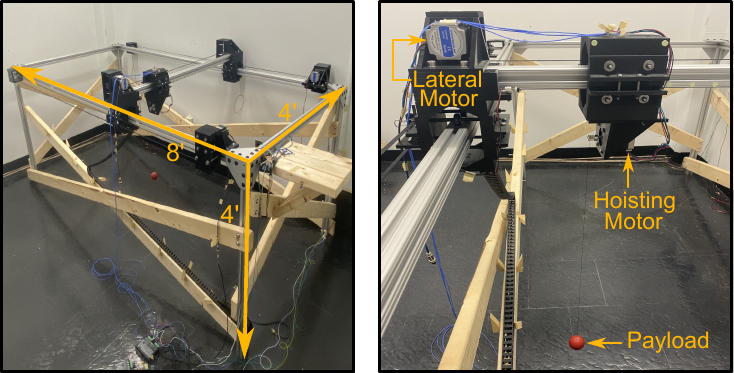}
\caption{Experimental Crane Setup. (a) Full View (b) Motor and Payload Zoom}\label{fig:crane}
\end{figure}

Experiments were conducted on an 8x4x4 foot (2.44x1.22x1.22 meter) scaled model of a gantry crane as seen in Fig. \ref{fig:crane}. The experimental data validates the spectrum of time-optimal control profiles and permits evaluating the residual energy in the system when compared to the rigid-body based control profiles. Multiple runs of the same control profile were conducted to generate statistics of the performance of the controller. Finally, a robust control profile was implemented and the variation in the residual energy as a function of errors in the initial length of the gantry cable was studied.

All of the experiments were conducted with the payload being hoisted down from its initial position, i.e., the final cable length is always greater than the initial cable length. We have conducted experiments with three unique control profiles, each of them with the same hoisting displacement to demonstrate the variation in the structure of the time-optimal profile analogous to that presented in Fig.~\ref{fig:switch_tf}. By fixing the hoisting displacement, we guarantee an evolution of the control profile as we modify the cart displacement. The first profile with the smallest cart displacement is a four switch profile. The second profile with a larger cart displacement is a two switch profile. The third profile with the greatest cart displacement is a three switch profile. The three switch profile is especially unique as it shows the optimal trajectory requires the cart motion to initially be quiescent while the hoisting is active. 

\begin{figure}
\centering
\includegraphics[trim=0 0 0 0,clip,width=\linewidth]{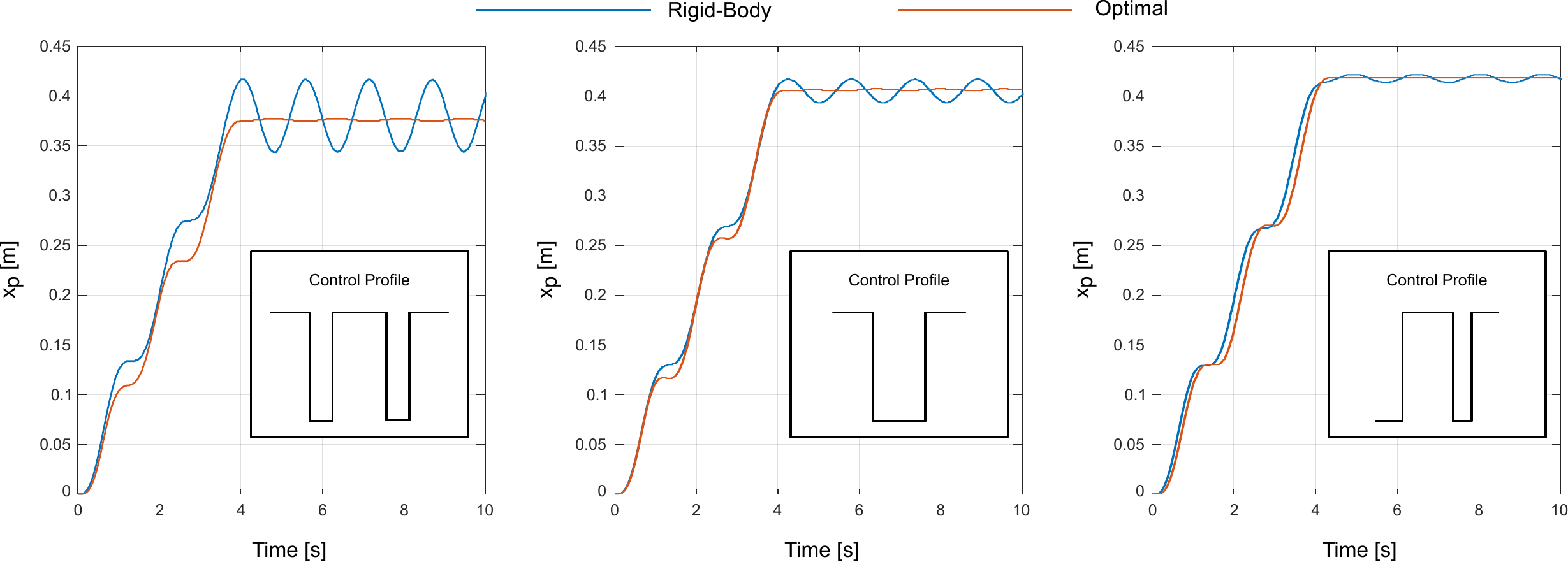}
\caption{Comparison of a Rigid-Body Maneuver to a Time-Optimal Maneuver. (a) Four Switch Profile with $x_f$ = 0.38 m (b) Two Switch Profile with $x_f$ = 0.41 m (c) Three Switch Profile with $x_f$ = 0.4242 m}\label{fig:Direct_Opt_Comp}
\end{figure}

For comparison for each of these optimal trajectories, we have also captured data of a rigid-body based controller. These controllers have the same final hoisting and cart displacements, and require the cart and hoisting motors to be driven at constant velocity over the duration of the maneuver. Fig.~\ref{fig:Direct_Opt_Comp} compares the performance of the rest-to-rest maneuvers of the time-optimal control profiles 
relative to the rigid-body based design. The figure shows the results for a four switch profile, two switch profile, and three switch profile, where the rigid-body maneuver times are 3.80 s, 4.10 s, and 4.24 s, respectively. The time-optimal maneuver times are 4.01 s, 4.17 s, and 4.40 s, respectively. We can see a significant reduction in the post-maneuver oscillations for the time-optimal profiles given a small penalty in the final maneuver time. A publicly available video analysis tool called Tracker~\cite{trackersoft} was used to extract position and velocity data of the suspended mass, after establishing a coordinate system and scale. Calibrating the software to a stationary object gives us a mean error of 0.163 mm. This calibration error or small discrepancies in the scale definition can be attributed to minor errors in the extracted data. 

A batch of experiments were collected for the rigid-body trajectory and optimal trajectory for the first optimal control profile with four switches. The data from these tests was used to show the repeatability of the experiments as well as to show the comparison of the residual mechanical energies. Ten experiments were conducted for each of the rigid-body and optimal four switch trajectories. The mechanical energy was calculated by isolating two complete oscillations after the completion of the maneuver. Using Eq.~\eqref{eq:Energy_mech}, 

\begin{equation}
E = \frac{1}{2}mv^2 + mgh \label{eq:Energy_mech}
\end{equation}
the mechanical energy at each time instant was evaluated and averaged across the two periods.

\begin{figure}[htbp]
\centering

\begin{minipage}[c]{0.45\textwidth}
    \centering
    \includegraphics[width=0.95\linewidth]{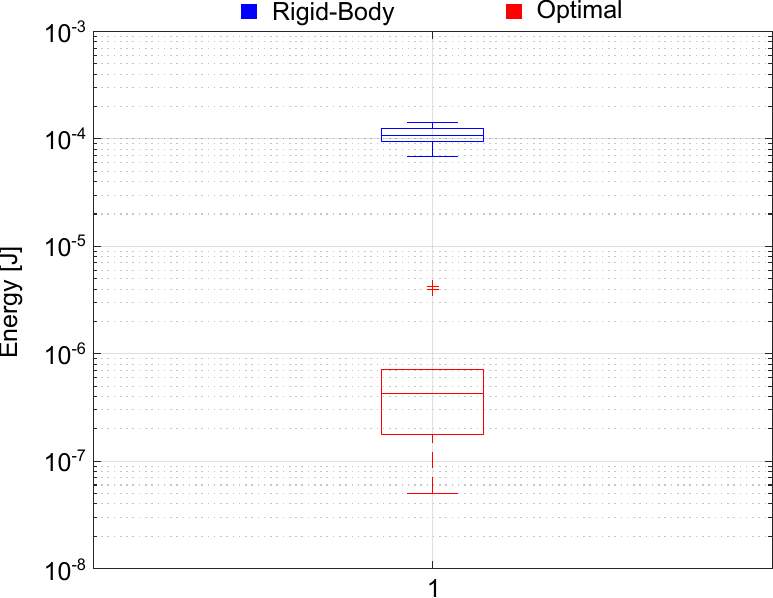}
    \captionof{figure}{Box and Whisker Energy Comparison}
    \label{fig:Box_Whisk}
\end{minipage}
\hfill
\begin{minipage}[c]{0.45\textwidth}
    \centering
    \scriptsize
    \begin{tabular}{lll}
        \toprule
        Experiment & Mean Energy & Variance \\
        \midrule
        Rigid-Body & 105.91 mJ & 558.22 mJ\textsuperscript{2} \\
        Optimal    & 1.08 mJ   & 2.55 mJ\textsuperscript{2}   \\
        \bottomrule
    \end{tabular}
    \captionof{table}{Energy Statistics}
    \label{tb:Energy}
\end{minipage}

\end{figure}

The box and whisker plot comparing these results is presented in Fig.~\ref{fig:Box_Whisk}. We note that this plot is on a logarithmic scale, which further highlights the profound improvement in energy reduction. Using the averages from all ten rigid-body experiments and all ten optimal experiments, there is a 98.98\% reduction in residual energy. The values for the average residual mechanical energy for each batch of experiments can be seen in Table~\ref{tb:Energy}. 


\begin{figure}
\centering
\includegraphics[width=.85\linewidth]{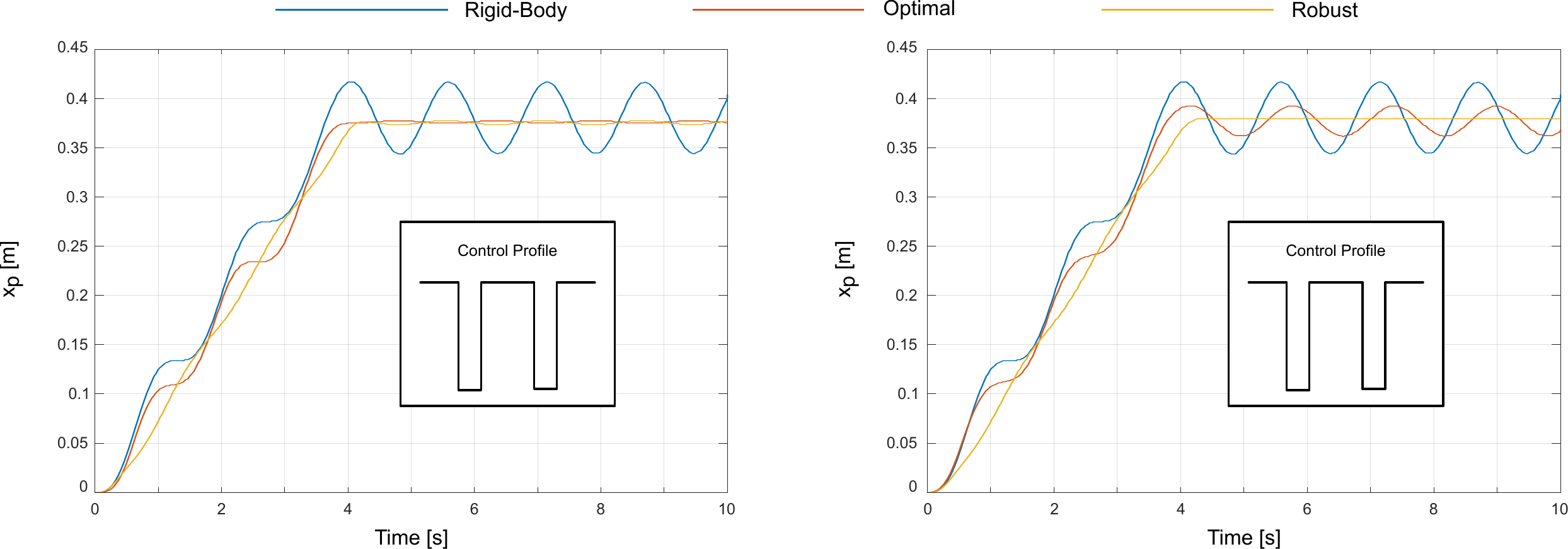}
\caption{Comparison of a Rigid-Body Maneuver, Time-Optimal Maneuver, and Robust Time-Optimal Maneuver. (a) Four Switch Profile with the Proper Initial Length (b) Four Switch Profile with an Initial Length Offset by 5 cm}\label{fig:Direct_Opt_Rob_Comp}
\end{figure}

Fig.~\ref{fig:Direct_Opt_Rob_Comp} compares the rigid-body and time-optimal trajectories to the robust time-optimal trajectories for the same terminal displacement and hoisting maneuver. The first figure illustrates the suspended mass displacement over time for a profile designed to reach a final displacement of 0.38 meters, with an initial cable length of 0.4 meters and a final cable length of 0.6 meters. The second figure displays the impact of an initial error in the suspended mass of 5 cm on the performance of the three controllers.  Fig.~\ref{fig:Direct_Opt_Rob_Comp}b clearly illustrates the sensitivity of the time-optimal solution manifested by an increase in the residual vibration, while the suspended mass is nearly stationary at the end of the maneuver for the robust controller, despite the error in the cable length $L_0$. The maneuver times for the rigid-body, optimal, and robust experiments in this figure are 3.80 s, 4.01 s, and 4.35 s, respectively. This shows that for a small penalty in maneuver time, we get significant energy reduction. 

\begin{figure}
\centering
\includegraphics[width=.4\linewidth]{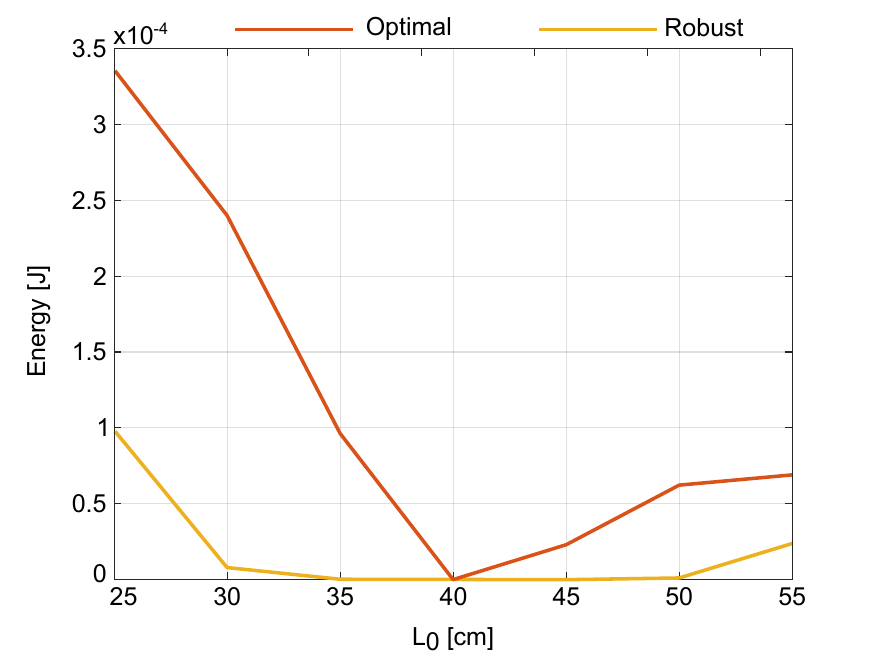}
\caption{Comparison of Residual Energy Between a Time-Optimal Design and a Robust Time-Optimal Design} \label{fig:Offset_Energy_Curves}
\end{figure}

To further analyze the performance of the robust time-optimal profile compared to the standard time-optimal profile, multiple experiments were conducted where the initial length is offset from the nominal length. The nominal initial length of the pendulum for every experiment is 40 cm. We will offset the initial length by $\pm 5$, $\pm 10$, and $\pm 15$ cm, resulting in seven distinct experiments. The residual mechanical energy of each experiment was calculated and presented in Fig. \ref{fig:Offset_Energy_Curves} which illustrates the reduced range of residual energy over the range of initial condition errors for the robust time-optimal controller.

\section{Conclusions}
The focus of this paper is on the development of time-optimal control profiles for the rest-to-rest maneuver of a gantry crane which includes hoisting. Assuming that the hoisting maneuver is completed in the same time as the cart completes its displacement results in a linear time-varying model for the gantry crane system. The optimal motion of the pendulum mass is shown to be parameterized by Bessel functions which permit transforming the time-optimal control problem to a parameter optimization problem. A comprehensive analysis of the impact of the cart displacement on the structure of the control profile is conducted revealing that the time-optimal control profile can be parameterized as a 2, 3, 4, or 6 switch profile over the domain of considered displacements. Constraints to determining the exact cart displacement which corresponds to the introduction or decimation of switches is presented. Sensitivity states are introduced to permit design of time-optimal control profiles with additional constraints to force the sensitivity of the states at the terminal time to zero. The resulting control profiles are shown to tradeoff performance (maneuver time) to robustness to initial condition uncertainties. Finally, numerous experiments are conducted to validate the numerical results presented in the paper. 

\appendix
\section{Reduction to Bessel's Equation}
\label{appendix_A}

The dynamics of the pendulum crane model considered in this paper can be abstracted to the linear time-varying system:
\begin{equation}
\frac{d^2x}{dt^2} + \frac{a}{b+ct} \, x(t) = \frac{a t}{b+ct},
\label{eq:main}
\end{equation}
where $a,b$, and $c$ are constants.

\subsection*{Homogeneous Solution}
To determine the homogeneous solution, consider the equation:
\begin{equation}
\frac{d^2x}{dt^2} + \frac{a}{b+ct} \, x = 0.
\end{equation}
The change of variable:
\[
s = b+ct, \qquad \frac{d}{dt} = c \frac{d}{ds}, \qquad \frac{d^2}{dt^2} = c^2 \frac{d^2}{ds^2}.
\]
permitting rewriting the equation as:
\begin{equation}
c^2 \frac{d^2x}{ds^2} + \frac{a}{s} x = 0
\quad \Longrightarrow \quad \frac{d^2x}{ds^2} + \frac{k}{s}x = 0,
\label{eq:sform}
\end{equation}
with $k = a/c^2$.

A second transformation $s=z^2$ leads to:
\[
\frac{dx}{ds} = \frac{1}{2z}\frac{dx}{dz}, \qquad
\frac{d^2x}{ds^2} = \frac{1}{4z^2}\frac{d^2x}{dz^2} - \frac{1}{4z^3}\frac{dx}{dz}.
\]
Substituting into Eq. \eqref{eq:sform}:
\begin{equation}
\frac{1}{4z^2}\frac{d^2x}{dz^2} - \frac{1}{4z^3}\frac{dx}{dz} + \frac{k}{z^2}x = 0.
\end{equation}
Multiplying through by $4z^2$:
\begin{equation}
\frac{d^2x}{dz^2} - \frac{1}{z}\frac{dx}{dz} + 4k x = 0.
\label{eq:Xeq}
\end{equation}
To eliminate the first-derivative term, consider the mapping $x(z)=z w(z)$. Then
\[
\frac{dx}{dz} = w+z\frac{dw}{dz}, \qquad \frac{d^2x}{dz^2} = 2\frac{dw}{dz} + z\frac{d^2w}{dz^2}.
\]
Substituting into Eq. \eqref{eq:Xeq}:
\begin{align}
(2\frac{dw}{dz}+z\frac{d^2w}{dz^2}) - \frac{1}{z}(w+z\frac{dw}{dz}) + 4kzw &= 0, \\
z^2 \frac{d^2w}{dz^2} + z \frac{dw}{dz} + (4k z^2 - 1) w &= 0 \label{eq:inteq}.
\end{align}

The final transformation to represent the system model in the standard Bessel equation form is:
\[
u = 2\sqrt{k} \, z = \frac{2\sqrt{a}}{|c|}\sqrt{b+ct}.
\]
which transforms Eq.~\eqref{eq:inteq} to:
\begin{equation}
u^2 \frac{d^2w}{du^2} + u \frac{dw}{du} + (u^2 - 1)w = 0,
\end{equation}
which is the standard Bessel equation of order $1$, whose closed form solution is:
\[
w(u) = C_1 J_1(u) + C_2 Y_1(u).
\]
Undoing the substitutions results in the final solution,
\begin{equation}
x_h(t) = \sqrt{b+ct}\Big( C_1 J_1\!\left(\tfrac{2\sqrt{a}}{|c|}\sqrt{b+ct}\right) + C_2 Y_1\!\left(\tfrac{2\sqrt{a}}{|c|}\sqrt{b+ct}\right) \Big).
\end{equation}

\subsection*{Particular Solution}
The RHS of Eq. \eqref{eq:main} is $a t/(b+ct)$. A simple trial solution $x_p(t)=t$ satisfies the equation:
\[
\frac{d^2x_p}{dt^2}=0, \qquad \frac{a}{b+ct}x_p(t) = \frac{a t}{b+ct}.
\]
Hence $x_p(t)=t$ is a valid particular solution.

Combining the homogeneous and particular parts leads to the general solution:
\begin{equation}
x(t) = t + \sqrt{b+ct}\Big( C_1 J_1\!\left(\tfrac{2\sqrt{a}}{|c|}\sqrt{b+ct}\right) + C_2 Y_1\!\left(\tfrac{2\sqrt{a}}{|c|}\sqrt{b+ct}\right) \Big).
\end{equation}




\section{Derivation of $\lambda_3$}
\label{appendix_B}

Let us substitute the standard form used in \ref{appendix_A} into Eqs. \eqref{eq:lam3} and \eqref{eq:lam2sol}:

\begin{align}
\dot{\lambda}_3 &=  -\lambda_2\frac{a}{b + ct} \label{eq:B1}\\
\lambda_2 &=  \sqrt{b + ct} \left(C_1 J_1 \left(\Phi(t) \right) + C_2 Y_1 \left(\Phi(t) \right)\right), \label{eq:B2}
\end{align}

where $\Phi(t) = 2\sqrt{k(b+ct)}$. Differentiating $\Phi(t)$ with respect to time:

\begin{equation}
\frac{d\Phi}{dt} = \frac{c\sqrt{k}}{\sqrt{b+ct}} = \sqrt{\frac{a}{b+ct}}\operatorname{sgn}(c). \label{eq:B3}
\end{equation}

Plugging Eq. \eqref{eq:B2} into Eq. \eqref{eq:B1}:

\begin{equation}
\dot{\lambda}_3 =  -\frac{a\sqrt{b+ct}(C_1J_1(2\sqrt{k(b+ct)}) + C_2Y_1(2\sqrt{k(b+ct)}))}{b + ct}. \label{eq:B4}
\end{equation}

Substitute Eq. \eqref{eq:B3} into Eq. \eqref{eq:B4} and integrating both sides:

\begin{equation}
\begin{aligned}
\int\dot{\lambda_3} \: dt &= -\frac{a\sqrt{b+ct}}{b+ct}\int(C_1 J_1(2\sqrt{k(b+ct)}) + C_2 Y_1(2\sqrt{k(b+ct)}))dt\\
&= -\sqrt{a}\operatorname{sgn}(c)\int(C_1 J_1(2\sqrt{k(b+ct)}) + C_2 Y_1(2\sqrt{k(b+ct)}))d\Phi.
\end{aligned}
\end{equation}

This provides us with the solution:

\begin{equation}
\lambda_3 = \sqrt{a}\operatorname{sgn}(c)(C_1 J_0(2\sqrt{k(b+ct)}) + C_2 Y_0(2\sqrt{k(b+ct)})) + C_3.
\end{equation}



\section*{Acknowledgment}
The authors acknowledge the support of this work by the US National Science Foundation through CMMI Award number 2021710. Author Eric Mountain acknowledges the support of the Prentice Family Foundation.

\vspace{-0.1in}
\bibliographystyle{elsarticle-num}
\bibliography{ref} 
\noindent\rule[0.5ex]{\linewidth}{1pt}

\newpage

\onecolumn


\end{document}